\documentclass[conference]{IEEEtran}
\usepackage{algorithmic}
\usepackage{graphicx}
\usepackage{textcomp}
\usepackage{xcolor}
\usepackage{caption}
\usepackage{subfigure}
\usepackage{algorithmic}
\usepackage{graphicx}
\usepackage{afterpage}
\usepackage{textcomp}
\usepackage{xcolor}
\usepackage{mathtools}
\usepackage{multirow}
\usepackage[many]{tcolorbox}
\usepackage{booktabs}
\usepackage{csvsimple}
\usepackage{longtable}
\usepackage{filecontents}
\usepackage{tabularx}
\usepackage{svg}
\usepackage{datatool}
\usepackage{adjustbox}
\usepackage{array}
\usepackage{xtab,booktabs}
\usepackage{tikz}
\usetikzlibrary{fit,arrows,calc,positioning}
\usepackage{color}
\usepackage{colortbl}
\usepackage{amsmath}

\usepackage{hyperref}
\usepackage{amssymb}
\usepackage{amsfonts}
\usepackage{makecell}
\usepackage{cite}
\usepackage{amsmath,amssymb,amsfonts}
\usepackage{algorithmic}
\usepackage{graphicx}
\usepackage{textcomp}
\usepackage{xcolor}
\def\BibTeX{{\rm B\kern-.05em{\sc i\kern-.025em b}\kern-.08em
    T\kern-.1667em\lower.7ex\hbox{E}\kern-.125emX}}
\newcommand{\ProjectName}{REARRANGE}
\newcommand{\new}[1]{{\color{black}{#1}}}

\begin{document}
\title{Closing the Loop for Software Remodularisation - \ProjectName: An Effort Estimation Approach for Software Clustering-based Remodularisation}




\author{\IEEEauthorblockN{Alvin Jian Jia Tan}
\IEEEauthorblockA{\textit{School of Information Technology} \\
\textit{Monash University Malaysia}\\
Malaysia \\
alvin.tan@monash.edu}
\and
\IEEEauthorblockN{Chun Yong Chong}
\IEEEauthorblockA{\textit{School of Information Technology} \\
\textit{Monash University Malaysia}\\
Malaysia \\
chong.chunyong@monash.edu}
\and
\IEEEauthorblockN{Aldeida Aleti}
\IEEEauthorblockA{\textit{Faculty of Information Technology} \\
\textit{Monash University}\\
Australia \\
aldeida.aleti@monash.edu}
}

\maketitle

\begin{abstract}
Software remodularization through clustering is a common practice to improve internal software quality. However, the true benefit of software clustering is only realized if developers follow through with the recommended refactoring suggestions, which can be complex and time-consuming. Simply producing clustering results is not enough to realize the benefits of remodularization. For the recommended refactoring operations to have an impact, developers must follow through with them. However, this is often a difficult task due to certain refactoring operations' complexity and time-consuming nature. 
\end{abstract}


\section{Introduction}
\new{In the work by Martini \cite{martini2016estimating}, the authors discussed that when 42 developer work months (DWM) were spent on refactoring, the effort spent on maintenance was reduced by 53.34 DWM, demonstrating a quantifiable benefit of refactoring. Ensuring high modularity pays off in the long term (from the perspective of development and maintenance effort), quantifying the business value of modularization. However, developers might be hesitant to follow the suggested refactoring operations due to the uncertain amount of effort needed, especially when the suggestions are ``Big Bang" remodularisation (refactoring operations that reorganize most of the system's classes into packages).} This work addresses this issue by introducing an approach that closes the loop in existing software clustering and remodularization research \cite{chong2017automatic,tan2022sc4r, chong2015constrained}. The proposed approach estimates the time required to carry out suggested refactoring operations based on the past evolutionary history of the analyzed software. By informing developers of potentially complicated and time-consuming refactoring operations, the approach helps developers prioritize refactoring efforts and plan accordingly. This allows practitioners to make more informed decisions about refactoring operations within time and budget constraints.

\section{Proposed Approach}
\begin{figure*}[!ht]
  \centering
    \includegraphics[scale=0.075]{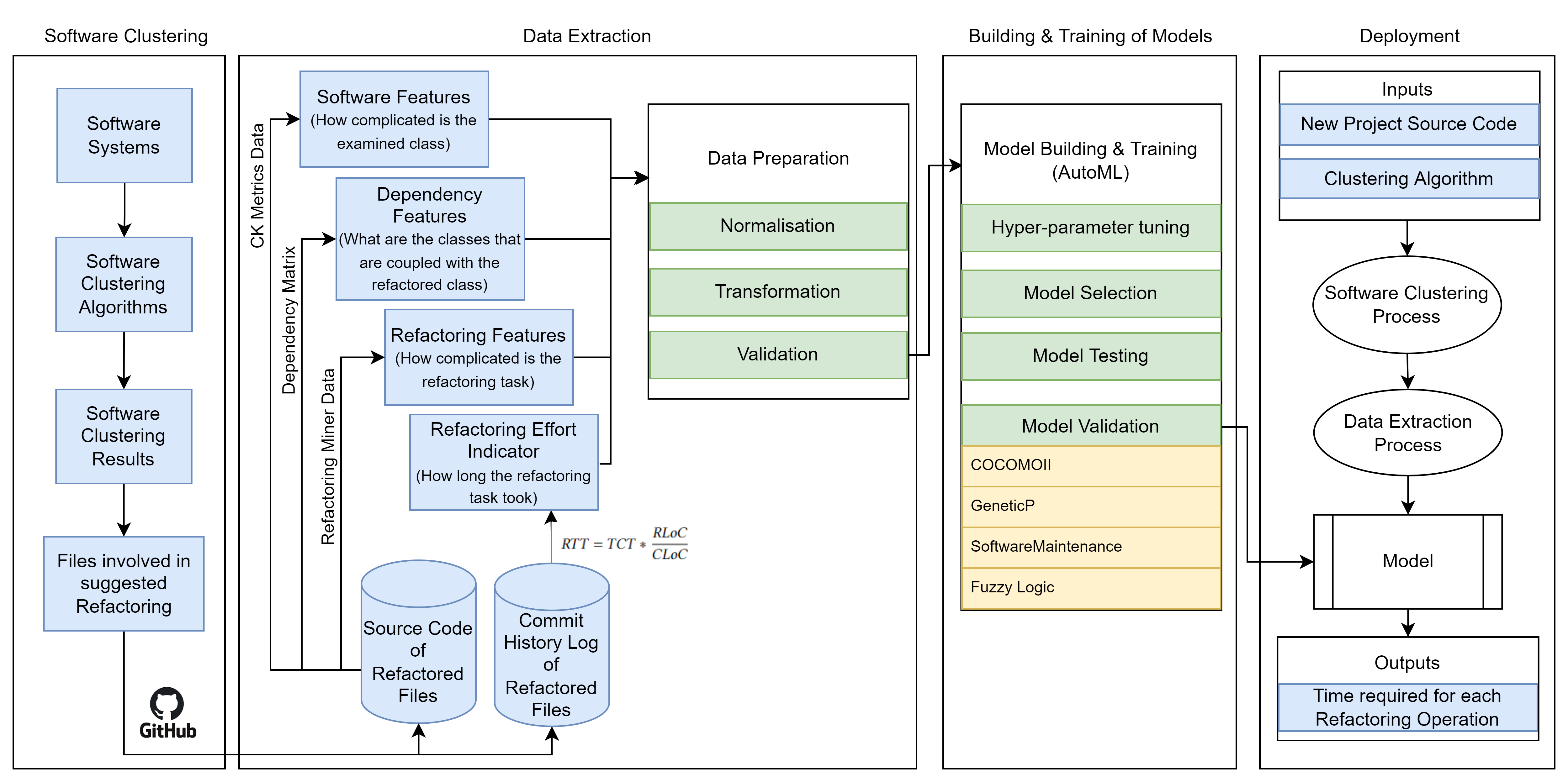}
    \caption{Overall workflow of \ProjectName.}
  \label{framework}
\end{figure*}

The proposed approach tries to \textbf{predict the effort required in person-hours to carry out the suggested refactoring operations given by the software clustering results} based on past refactoring operations performed by the developers, which involves four main parts, depicted in Figure \ref{framework}. 


\subsubsection{Software Systems}
\new{The software systems\footnote{\url{https://figshare.com/s/0d24fa70b21104e7dcf1}} used in this study are chosen pseudo-randomly, with a mixture of GitHub projects and Apache projects to provide a meaningful representation of different types of open-source projects with various sizes and complexities.}

\subsubsection{Refactoring Effort Indicator}
\new{A proxy effort is then generated to estimate the refactoring effort \cite{schroder2021search}. Taking Apache Camel project and one of its commits as an example, Total Commit Time ($TCT$) = $9$ person-hours, Refactored Lines of Code ($RLoC$) = $25$ lines, Commit Lines of Code ($CLoC$) = $102$ lines. The Refactoring Time Taken ($RTT$) for this commit is given as $9\times \frac{25}{102}$ = 2.205 person-hours.}


\subsubsection{Data Extraction of Features}
\textbf{Software Features - }\new{for each refactoring operation in the analyzed project, the software features of the classes involved in the refactoring operation are extracted using the Java CK (Chidamber and Kemerer) code metrics calculator \cite{ CKmetric}.} \textbf{Refactoring Features -} \new{we use Refactoring Miner \cite{Tsantalis:TSE:2020:RefactoringMiner2.0} to detect refactoring operations applied throughout the evolution of a project. Refactoring Miner records all refactoring operations between each minor version of the project, from the first to the last release.} \textbf{Dependency Features - }\new{we employ \textit{Depends} \cite{jin2019enre} to extract dependencies between classes in the analyzed software. \textit{Depends} gathers syntactical relations among source code entities, such as files and methods, and extracts different types of dependencies from a pair of classes, such as import, contain, call, return, and implement.}

\subsubsection{Training the Machine Learning Models}
\new{H2O AutoML~\cite{H2OAutoML20} is used to learn the relationship between features and the proxy effort. In our work, the Gradient Boosting Machine method performs the best as a regression model to predict refactoring effort.}

\section{Results}
The MAE (Mean Absolute Error) of 5.47 for \ProjectName~is within an acceptable error of 1 working day of 8 person-hours. This step answers the first research question, \textit{RQ1: How effective is REARRANGE in predicting the effort required for implementing the clustering results?} By comparing \ProjectName~against sanity checks, we answer the second research question, \textit{RQ2: Does \ProjectName~satisfy the requirements of a baseline estimation model?} Finally, we compare \ProjectName~against other software cost estimation models  as shown in Table~\ref{benchmark_regression}. By comparing MAE between \ProjectName~(5.47) and Genetic Estimation (453.31) as the next best alternative from Table \ref{benchmark_regression}, it is evident that \ProjectName~performs better at estimating refactoring effort. The significant difference in MAE is because we compare \ProjectName~to a software effort estimation model. The lack of a direct comparison further highlights the need for a refactoring effort-focused estimation model. The RMSE results show that the existing software cost estimation models are not good estimators of the refactoring effort, answering  the third research question, \textit{RQ3: How does \ProjectName~compare to other software effort estimation models?} 

\begin{table}[h!]
\begin{center}
\caption{Benchmark for \ProjectName~(person-hour)}
\begin{tabular}{|c|c|c|c|} 
 \hline
 \textbf{Estimation Model} & \textbf{R2} & \textbf{RMSE} & \textbf{MAE} \\ 
 \hline
 Mean & -0.974 & 21.37 & 18.93 \\ 
 \hline
 \textbf{\ProjectName} & \textbf{0.538} & \textbf{15.38} & \textbf{5.47}\\ 
 \hline
 COCOMOII & -116903 & 5201.57 & 893.22 \\ 
 \hline
 GeneticP & -21469.5 & 2229.15 & 453.31\\ 
 \hline
\end{tabular}
\label{benchmark_regression}
\end{center}
\end{table}
\vspace{-1em}

Although \ProjectName~may sometimes overestimate the time required for smaller refactoring operations, it can accurately identify and predict large and complex refactoring operations that are incredibly disruptive to developers. Similarly, existing software estimation models significantly overestimate the time required compared to \ProjectName, resulting in high RMSE and MAE values. Developer feedback was then obtained on the proposed refactoring operations and the estimated efforts of the chosen test subjects. This is done by filing issue reports on the relevant GitHub repository or Jira. Among the total 53 suggested refactoring operations and its estimated effort required, 23 received a positive response with only one negative response. Overall, the positive sentiments from developers show that \ProjectName~is able to provide reasonable estimates that are relevant to software developers. Table \ref{developer_validation} shows sample responses from developers.


\begin{table}[h!]
\begin{center}
\caption{\new{Results from Developer Validation}}
\begin{tabular}{|c|c|c|c|c|c|} 
\hline
 \makecell{\textbf{Project}\\ \textbf{Name}} & \makecell{\textbf{Issue}\\ \textbf{ID}} & \makecell{\textbf{No. of } \\ \textbf{Refactoring} \\ } & \textbf{Positive} & \textbf{Negative} & \textbf{Status} \\
 \hline
 Dropwizard & \href{https://github.com/dropwizard/dropwizard/discussions/5402}{5402} & 4 & 0 & 0 & Pending \\ 
 \hline
 Redisson & \href{https://github.com/redisson/redisson/issues/4363}{4363} & 5 & 0 & 0 & Pending \\ 
\hline
 Okhttp & \href{https://github.com/square/okhttp/issues/7355}{7355} & 17 & 17 & 0 & Closed \\
 \hline
 Java-tron & \href{https://github.com/tronprotocol/java-tron/issues/4500}{4500} & 2 & 1 & 1 & Closed \\
 \hline
\end{tabular}
\label{developer_validation}
\end{center}
\end{table}
\vspace{-2em}

\small{
\bibliographystyle{IEEEtran}
\bibliography{references}
}
\end{document}